\documentclass{article}
\usepackage[T1]{fontenc} 
\usepackage[utf8]{inputenc} 
\usepackage{ismir,amsmath,cite,url}
\usepackage{graphicx}
\usepackage{color}
\usepackage{booktabs}
\usepackage{lineno}
\usepackage[bookmarks=false]{hyperref}
\hypersetup{colorlinks=true,urlcolor=blue}

\title{Exploring Internet Radio Across the Globe with the MIRAGE Online Dashboard}

\fourauthors
  {Ngan V.T. Nguyen} {University of Science, VNU-HCMUS \\ {\tt \small nvtngan@hcmus.edu.vn}}
  {Elizabeth A.M. Acosta} {Texas Tech University\\{\tt \small liz.acosta@ttu.edu}} 
  {Tommy Dang} {Texas Tech University \\ {\tt \small tommy.dang@ttu.edu}}
  {David R.W. Sears} {Texas Tech University\\{\tt \small david.sears@ttu.edu}}
  
\def\authorname{Nguyen et al.}

\begin{document}

\maketitle
\begin{abstract}
This study presents the \textit{Music Informatics for Radio Across the GlobE} (MIRAGE) online dashboard, which allows users to access, interact with, and export metadata (e.g., artist name, track title) and musicological features (e.g., instrument list, voice type, key/mode) for 1 million events streaming on 10,000 internet radio stations across the globe. Users can search for stations or events according to several criteria, display, analyze, and listen to the selected station/event lists using interactive visualizations that include embedded links to streaming services, and finally export relevant metadata and visualizations for further study. 
\end{abstract}

\section{Introduction}\label{sec:introduction}

Despite its scholarly neglect relative to television, film, and print \cite{Lacey2018}, radio’s convergence with the internet has extended its reach via web browsers and smartphone apps, enabling the medium to persist as a central site of culture and daily life for communities around the world \cite{Bottomley2020, Glantz2016}. The recent resurgence of pirate and community radio stations on the internet alongside national and multinational networks also reflects internet radio’s lower production costs relative to short-wave terrestrial (e.g., FM or AM) radio \cite{Wall2004}, resulting in a diverse range of both standardized and specialized programming \cite{Chambers2003, Hendy2000, Uimonen2017}. 

And yet, the volume and scope of much of the research in fields like radio studies has been freighted heavily towards the Global North \cite{Lacey2018}. 
In doing so, the research program just described attempts to situate listeners within a particular musical tradition (e.g., western classical or popular music), rather than within a particular geographic environment (e.g., El Paso, Texas) where myriad musical traditions might co-exist. As a result, music's vast global marketplace has yet to receive sustained scholarly attention in the MIR community. 

To address this issue, this study presents the development release (v0.2) of the \textit{Music Informatics for Radio Across the GlobE} (MIRAGE) online dashboard, which allows users with potentially little training in computational methods to access, interact with, and export metadata (e.g., artist name, track title) and musicological features (e.g., instrument list, voice type, key/mode) for 1 million events streaming on 10,000 internet radio stations across the globe. To that end, Section~\ref{sec:background} summarizes previous research on the development of digitized music corpora and cultural databases. Next, Section~\ref{sec:metacorpus} presents the MIRAGE-MetaCorpus, Section~\ref{sec:dashboard} introduces the MIRAGE online dashboard, and Section~\ref{sec: use case} offers a potential use case. Finally, Section~\ref{sec:conclusion} discusses limitations and future directions for the MIRAGE project.  

\section{Previous Research}\label{sec:background}

In recent years, researchers in music theory, music information retrieval (MIR), and radio/media studies have developed digitized music corpora and cultural databases that represent data in machine-readable symbolic and audio formats. 

\begin{figure*}[t!]
    \centering
    \begin{minipage}[t!]{.45\linewidth} 
        \centering
        \begin{tabular}{rrrrr} 
        \toprule
              & \multicolumn{2}{c}{\textbf{Radio}} & \multicolumn{2}{c}{\textbf{Sovereign}} \\
        \multicolumn{1}{l}{\textbf{Continent}} & \multicolumn{2}{c}{\textbf{Stations}} & \multicolumn{2}{c}{\textbf{States}} \\
        \midrule
        \multicolumn{1}{l}{Africa} &        392  & \multicolumn{1}{l}{(4\%)} & 39    & \multicolumn{1}{l}{(58\%)} \\
        \multicolumn{1}{l}{Asia} &        653  & \multicolumn{1}{l}{(7\%)} & 38    & \multicolumn{1}{l}{(59\%)} \\
        \multicolumn{1}{l}{Europe} &     5,161  & \multicolumn{1}{l}{(52\%)} & 49    & \multicolumn{1}{l}{(60\%)} \\
        \multicolumn{1}{l}{North America} &     2,243  & \multicolumn{1}{l}{(22\%)} & 33    & \multicolumn{1}{l}{(67\%)} \\
        \multicolumn{1}{l}{Oceania} &        222  & \multicolumn{1}{l}{(2\%)} & 5     & \multicolumn{1}{l}{(17\%)} \\
        \multicolumn{1}{l}{South America} &     1,329  & \multicolumn{1}{l}{(13\%)} & 13    & \multicolumn{1}{l}{(87\%)} \\
        \textbf{TOTAL} &   10,000  &       & 177   &  \\
        \bottomrule
        \end{tabular}%
    \end{minipage}
    \hspace{2em}
    \begin{minipage}[t!]{.45\linewidth}
        \centering
        \includegraphics[scale=.058]{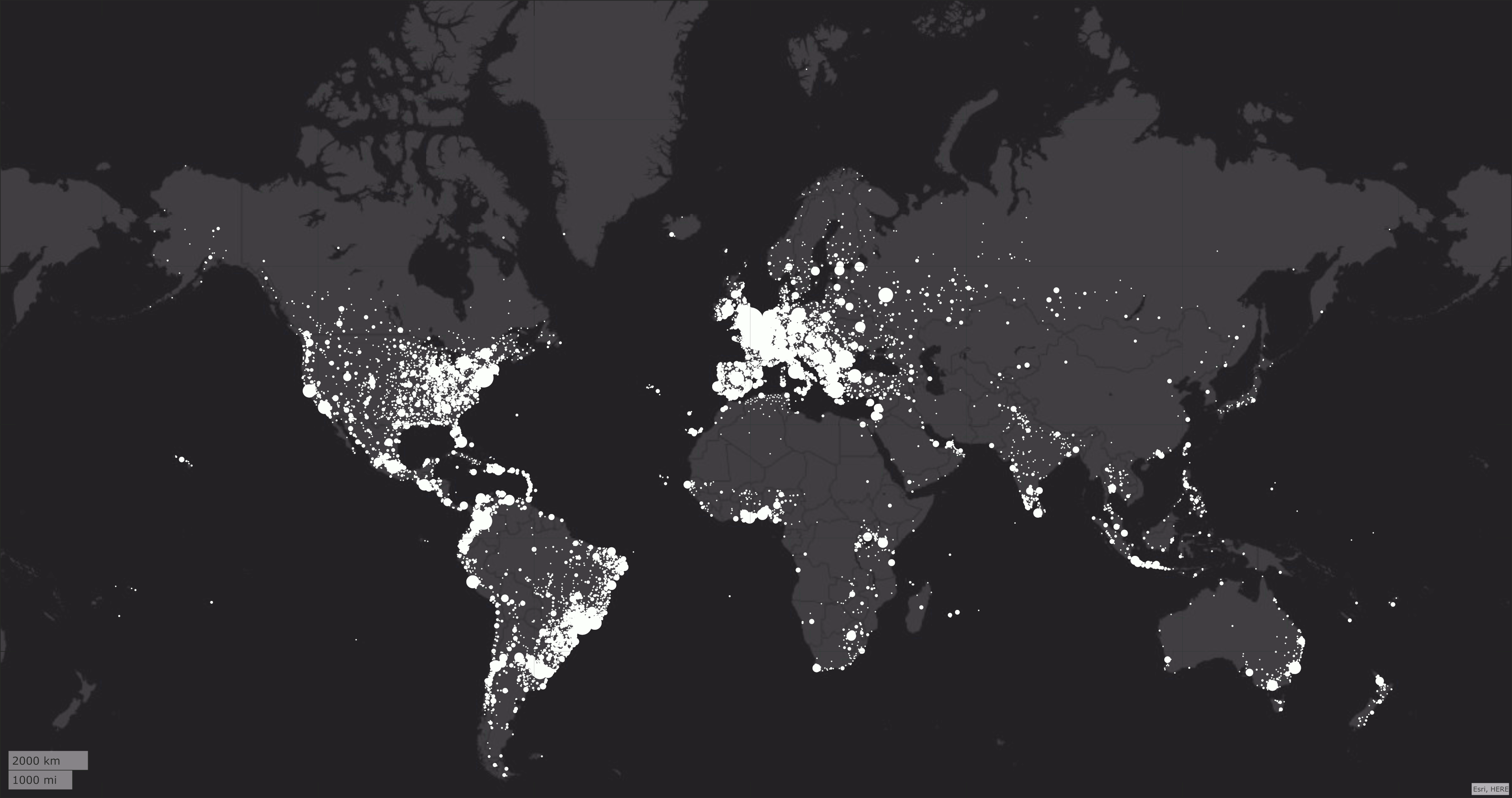}
    \end{minipage}
    \caption{Descriptive statistics (left) and geographic map (right) of the radio stations in MIRAGE-MetaCorpus. The size of each bubble represents the number of stations at that location.}
    \label{fig:tableANDmap}
\end{figure*}

In computational music theory, heavily curated corpora (100s of songs) like the McGill Billboard and Rolling Stone-200 data sets include expert annotations for musical parameters like harmony, meter, and melody, for example, but remain restricted to Anglophone popular music traditions \cite{Burgoyne2011,Declercq2011}. What is more, the limited size of symbolic corpora makes comparative research especially difficult \cite{Sears2021}. 

In MIR, corpora like the Million Song data set address issues of scale while avoiding copyright infringement by providing researchers with publicly available metadata and musicological features protected under fair use for a large collection of songs hosted on commercial music-streaming services like last.fm \cite{Bertin-Mahieux2011}. Nevertheless, the size, scope, and format of these projects require extensive training in distant-reading (i.e., computational) methods \cite{Aizenberg2012, Dang2012, Silva2017}. As a result, MIR corpora sometimes eschew the kinds of musical engagements favored by scholars in humanities disciplines using close-reading methodologies. Finally, the projects referenced above do not include information about the geographic location of the music encountered by listeners in everyday life.


Finally, in radio/media studies, researchers routinely employ interview and survey methodologies to explore radio stations across the globe \cite{AlaFossi2008, Lind1999}, in some cases by selecting samples from radio-station directories hosted online \cite{Kuhn2011}. The now defunct ComFM, for example, included a catalogue of web-radio stations classified according to geographic region and type of programming. Other current internet radio directories like radio.co and internet-radio.com offer searchable databases consisting of several thousand stations, but they do not permit users to access or export the entire database for further analysis. 

The MIRAGE online dashboard addresses these issues by offering a global archive of the musical traditions encountered on internet radio. For this reason, the dashboard's database could serve MIR tasks like music recommendation and genre classification, but the dashboard itself also allows researchers with potentially little training in computational methods to select and analyze a subset of events or stations (i.e., to develop their own sub-corpora). Finally, like previous MIR projects \cite{Bertin-Mahieux2011}, the MIRAGE online dashboard avoids copyright infringement by including publicly available metadata and musicological features protected under fair use while enabling users to stream recordings using embedded links to commercial services like Spotify and YouTube. 



\section{MIRAGE MetaCorpus}\label{sec:metacorpus}

The core database for the MIRAGE online dashboard is MIRAGE-MetaCorpus, which currently consists of metadata and musicological features for 1 million events that streamed on 10,000 internet radio stations across the globe. In this context, an `event’ could represent a musical work of some kind, or a radio program like a podcast or a call-in show. 

\subsection{Collecting MetaData}

Following \cite{Aizenberg2012}, data collection consisted of three stages: station-list and event-list collection (\textit{Stage 1}), station-list review (\textit{Stage 2}), and event-list parsing (\textit{Stage 3}).

\subsubsection{Stage 1: Collecting Station/Event Lists}
Toward Stage 1, the research team collected metadata for an initial list of internet radio stations and then monitored the station streams to obtain additional metadata from the stream encoder. To that end, we monitored radio stations in real time on Radio Garden,\footnote{\href{https://radio.garden}{https://radio.garden}.} a streaming service with an open-access application programming interface (API) that allows users to select and play publicly available radio streams using an interactive representation of the globe. 

Between the months October to January 2022-2023, a random sample of 10,000 stations from the initial station list was monitored throughout the 24-hour day -- but avoiding each ten-minute period at the top and bottom of the hour when advertising is most frequent -- in order to obtain additional metadata from the stream encoder for 100 events from each station, resulting in an initial list of 1 million events. The monitoring algorithm also excluded an event if the stream description did not include metadata, or if the metadata featured advertising terms or reflected a station blackout period (e.g., `advert', `commercial', `unknown', `blackout', etc.).

During event-list collection, additional metadata for each location in the initial station list was also included from the Natural Earth map data set,\footnote{\href{https://www.naturalearthdata.com/}{https://www.naturalearthdata.com/}.} which provides public-domain vector and map raster data along with accompanying metadata.  

Shown in Figure \ref{fig:tableANDmap}, the selected station list represents 177 of the globe's 305 sovereign states. As a random sample of Radio Garden's station list (i.e., the \textit{Radio Garden sample}), this release of the MIRAGE-MetaCorpus (v0.2) therefore reflects the prevalence of internet radio stations across the globe on the Radio Garden streaming service.

\begin{table}[t!]
  \centering
    \small
    \begin{tabular}{l|l}
        \toprule
        \multicolumn{1}{p{11.8em}|}{\hspace{-.5em} \textbf{\texttt{<location>}}} &  \\
        \multicolumn{1}{p{11.8em}|}{\hspace{-.5em} \enspace \texttt{<city>}\textsuperscript{ab}} & \multicolumn{1}{p{11em}}{ \hspace{-.5em} Johor Bahru} \\
        \multicolumn{1}{p{11.8em}|}{\hspace{-.5em} \enspace \texttt{<country>}\textsuperscript{ab}} & \multicolumn{1}{p{11em}}{\hspace{-.5em} Malaysia} \\
        \multicolumn{1}{p{11.8em}|}{\hspace{-.5em} \enspace \texttt{<country\_GDP>}\textsuperscript{b}} & \multicolumn{1}{p{11em}}{\hspace{-.5em} 863 Billion} \\
        \multicolumn{1}{p{11.8em}|}{\hspace{-.5em} \enspace \texttt{<coordinates>}\textsuperscript{ab}} & \multicolumn{1}{p{11em}}{\hspace{-.5em} 103.6545°, 1.4783°} \\
        \multicolumn{1}{p{11.8em}|}{\hspace{-.5em} \enspace \textbf{\texttt{<station>} }} &  \\
        \multicolumn{1}{p{11.8em}|}{\hspace{-.5em} \enspace \enspace \texttt{<name>}\textsuperscript{cd} } & \multicolumn{1}{p{11em}}{\hspace{-.5em} Best FM } \\
        \multicolumn{1}{p{11.8em}|}{\hspace{-.5em} \enspace \enspace \texttt{<form>}\textsuperscript{cd}} & \multicolumn{1}{p{11em}}{\hspace{-.5em} Simulcast (FM 104.1)} \\
        \multicolumn{1}{p{11.8em}|}{\hspace{-.5em} \enspace \enspace \texttt{<format>}\textsuperscript{d}} & \multicolumn{1}{p{11em}}{\hspace{-.5em} Adult Contemporary} \\
        \multicolumn{1}{p{11.8em}|}{\hspace{-.5em} \enspace \enspace \texttt{<genre>}\textsuperscript{cd}} & \multicolumn{1}{p{11em}}{\hspace{-.5em} pop, Indonesian pop} \\
        \multicolumn{1}{p{11.8em}|}{\hspace{-.5em} \enspace \enspace \texttt{<website>}\textsuperscript{cd}} & \multicolumn{1}{p{11em}}{\hspace{-.5em} \href{http://www.bestfm.com.my}{http://www.bestfm.com.my}} \\
        \multicolumn{1}{p{11.8em}|}{\hspace{-.5em} \enspace \enspace \textbf{\texttt{<event>}}} &  \\
        \multicolumn{1}{p{11.8em}|}{\hspace{-.5em} \enspace \enspace \enspace \texttt{<time@station>}\textsuperscript{c}} & \multicolumn{1}{l}{\hspace{-.5em} 12/28/2022 9:37} \\
        \multicolumn{1}{p{11.8em}|}{\hspace{-.5em} \enspace \enspace \enspace \texttt{<description>}\textsuperscript{c}} & \multicolumn{1}{p{11em}}{\hspace{-.5em} Aisha Retno – Sutera} \\
        \multicolumn{1}{p{11.8em}|}{\hspace{-.5em} \enspace \enspace \enspace \texttt{<reliability>}\textsuperscript{e}} & \multicolumn{1}{p{11em}}{\hspace{-.5em} 1} \\
        \multicolumn{1}{p{11.8em}|}{\hspace{-.5em} \enspace \enspace \enspace \textbf{\texttt{<artist>}}} &  \\
        \multicolumn{1}{p{11.8em}|}{\hspace{-.5em} \enspace \enspace \enspace \enspace \enspace \texttt{<name>}\textsuperscript{f}} & \multicolumn{1}{p{11em}}{\hspace{-.5em} Aisha Retno} \\
        \multicolumn{1}{p{11.8em}|}{\hspace{-.5em} \enspace \enspace \enspace \enspace \enspace \texttt{<type>}\textsuperscript{f}} & \multicolumn{1}{p{11em}}{\hspace{-.5em} musical artist} \\
        \multicolumn{1}{p{11.8em}|}{\hspace{-.5em} \enspace \enspace \enspace \enspace \enspace \texttt{<gender>}\textsuperscript{f}} & \multicolumn{1}{p{11em}}{\hspace{-.5em} female} \\
        \multicolumn{1}{p{11.8em}|}{\hspace{-.5em} \enspace \enspace \enspace \enspace \enspace \texttt{<country>}\textsuperscript{f}} & \multicolumn{1}{p{11em}}{\hspace{-.5em} Malaysia} \\
        \multicolumn{1}{p{11.8em}|}{\hspace{-.5em} \enspace \enspace \enspace \enspace \enspace \texttt{<genre>}\textsuperscript{f}} & \multicolumn{1}{p{11em}}{\hspace{-.5em} pop} \\
        \multicolumn{1}{p{11.8em}|}{\hspace{-.5em} \enspace \enspace \enspace \enspace \enspace \texttt{<instruments>}\textsuperscript{f}} & \multicolumn{1}{p{11em}}{\hspace{-.5em} piano, voice } \\
        \multicolumn{1}{p{11.8em}|}{\hspace{-.5em} \enspace \enspace \enspace \textbf{\texttt{<track>}}} &  \\
        \multicolumn{1}{p{11.8em}|}{\hspace{-.5em} \enspace \enspace \enspace \enspace \enspace \texttt{<title>}\textsuperscript{f}} & \multicolumn{1}{p{11em}}{\hspace{-.5em} Sutera} \\
        \multicolumn{1}{p{11.8em}|}{\hspace{-.5em} \enspace \enspace \enspace \enspace \enspace \texttt{<duration>}\textsuperscript{f}} & \multicolumn{1}{l}{\hspace{-.5em} 03:18} \\
        \multicolumn{1}{p{11.8em}|}{\hspace{-.5em} \enspace \enspace \enspace \enspace \enspace \texttt{<year\_released>}\textsuperscript{f}} & \multicolumn{1}{l}{\hspace{-.5em} 2022} \\
        \multicolumn{1}{p{11.8em}|}{\hspace{-.5em} \enspace \enspace \enspace \enspace \enspace \texttt{<key>}\textsuperscript{f}} & \multicolumn{1}{p{11em}}{\hspace{-.5em} C minor} \\
        \multicolumn{1}{p{11.8em}|}{\hspace{-.5em} \enspace \enspace \enspace \enspace \enspace \texttt{<language>}\textsuperscript{f}} & \multicolumn{1}{p{11em}}{\hspace{-.5em} Malay} \\
        \bottomrule
    \end{tabular}%
    \caption{Left: Selected variables from the encoding scheme for MIRAGE-MetaCorpus, expressed in pseudocode. Metadata were obtained from the following sources: \textsuperscript{a} Radio Garden API; \textsuperscript{b} Natural Earth map data set; \textsuperscript{c} Internet Radio Station Stream Encoder; \textsuperscript{d} Annotator Review; \textsuperscript{e} Monitoring/Matching Algorithm; \textsuperscript{f} Online Music Libraries. Right: An example of the metadata for an event in MIRAGE-MetaCorpus.}
  \label{tab:encoding}
\end{table}

\subsubsection{Stage 2: Reviewing Stations}
Toward Stage 2, a team of six human annotators began reviewing station-level metadata from the Radio Garden API and radio-station stream encoder in 2023-2024. For each station, an annotator reviewed the station's website url, station name, city, and country for incorrect/missing spelling, capitalization, punctuation, and diacritics. Next, the list of genres, formats, and terrestrial (FM/AM) station frequencies (if applicable) were reviewed and/or included using information on the station website. Finally, the annotator reviewed the corresponding event list for each station to determine the percentage of events that featured reliable stream-description metadata (i.e., artist name, track title). 

Currently, the research team has reviewed over 6,000 stations and plans to complete station-list review by 2025. 

\subsubsection{Stage 3: Parsing Events}

Toward Stage 3, additional metadata were collected for each event using the Spotify and WikiData online music libraries.\footnote{\href{https://open.spotify.com}{https://open.spotify.com}; \href{https://www.wikidata.org}{https://www.wikidata.org}.} Specifically, the team queried each API using each event's stream description. The obtained list of matching queries was then filtered using a normalized edit distance measure. Query lists featuring more than one matching entry based on normalized edit distance were then ranked by release date, and the track with the oldest release date was selected.



\begin{figure*}[t!]
  \centering
  \includegraphics[width=.8\textwidth,scale=.39]{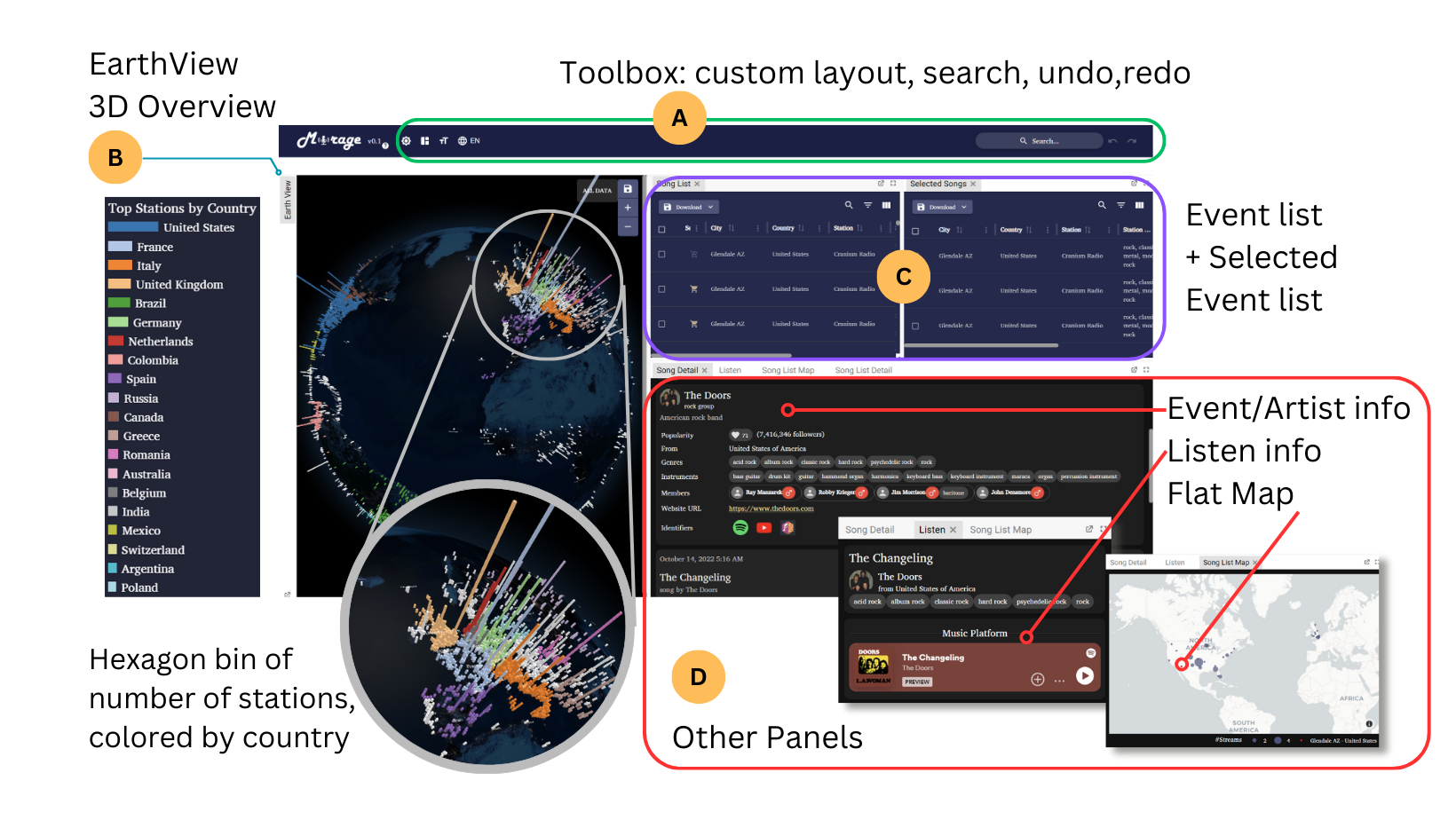}
  \caption{An overview of the MIRAGE online dashboard (v0.2).}
  \label{fig:mdash}
\end{figure*}

\subsection{MetaData Variables}


Each event in MIRAGE-MetaCorpus includes metadata for 100 variables obtained from the Radio Garden API (RG), the Natural Earth map data set (NE), the internet radio station stream encoder (SE), annotator review (AR), or using the online music libraries WikiData (WD), MusicBrainz (MB), Spotify (SP), Musixmatch (MX), YouTube (YT), Genius (GE), and AZlyrics (AZ). Shown in Table \ref{tab:encoding}, these metadata reflect information about each event's location, station, event, artist, and track. For example, location metadata includes variables like the city, country, and geographic coordinates of the monitored event, as well as demographic data like the country's population and GDP. Station metadata includes its name, form (a webcast stream, or a stream simulcast on the internet and terrestrial radio frequencies), formats (e.g., Top 40), and the station’s website url. Event metadata includes variables like the local time when the station was monitored and the event's identifying metadata, such as the name of the artist and title of the recording. Finally, artist and track metadata include variables like the name and type of the artist, and if the artist is a group, a list of the group’s members and their demographic information (their listed genders, sexual orientations, and ethnicities), the group’s country of origin by birth and/or citizenship, the title and duration of the track, and its year of release. 

\subsection{MetaData Access \& Export}

Users may access the complete MIRAGE-Metacorpus with the online dashboard.\footnote{The MIRAGE online dashboard is available at \href{https://pearl-laboratory.github.io/mirage-mc/}{https://pearl-laboratory.github.io/mirage-mc/}.} In addition, public-domain metadata from MIRAGE-MetaCorpus are available for download in an open-access repository on Zenodo \cite{Sears2024}, which includes both the complete data set and a subset of the data set for which the metadata obtained from the station's stream encoder and the corresponding metadata provided by online music libraries was deemed a reliable match (i.e., where the normalized edit distance measure between the two metadata character strings was $\geq$.90 on a 0–1 scale).  

\section{MIRAGE online dashboard}\label{sec:dashboard}

The MIRAGE online dashboard is an open-access web application that enables users to effortlessly navigate and engage with radio-station metadata and musicological features at various levels of detail. The dashboard's layout consists of fully interactive panes displaying relevant information from MIRAGE-MetaCorpus. The dashboard is also compatible with multiple platforms and operating systems, so users may access and interact with the dashboard from any internet-connected device. 


The complete technology stack of the dashboard includes Node.js for the server, MongoDB for the database, and React for the front end. This integration of technology guarantees a smooth user experience and effective data processing. Moreover, the dashboard may be tailored to accommodate individual users' distinct requirements and inclinations, rendering it a versatile instrument for analyzing radio stations. What is more, incorporating these technologies enables instantaneous data updates and interactive functionalities, thereby boosting the overall user experience over subsequent versions of the dashboard. In addition, the MIRAGE dashboard offers sophisticated search and filtering tools and the ability to export metadata and visualizations in URL, CSV, PNG, and SVG formats, allowing users to study and share data easily.  
 
\subsection{Structure \& Processing}\label{subsec:processing}

Shown in Figure~\ref{fig:mdash}, the MIRAGE dashboard's layout is divided into two groups: a toolbox on the top (A) and data-visualization panels below (B-D), making it easy for users to navigate and analyze information. The toolbox at the top includes options for language preference, panel-display customization, and searching. The data visualization panels show the data in various formats, such as charts, graphs, and tables, for straightforward interpretation and analysis. The panels can also dock to allow the user to create a customized layout, or open to another window (or undock) to permit a more detailed view suitable for multiple-screen presentations.

\begin{figure}[b!]
  \centering
  \begin{minipage}{1\linewidth} 
    \includegraphics[width=\textwidth,scale=.39]{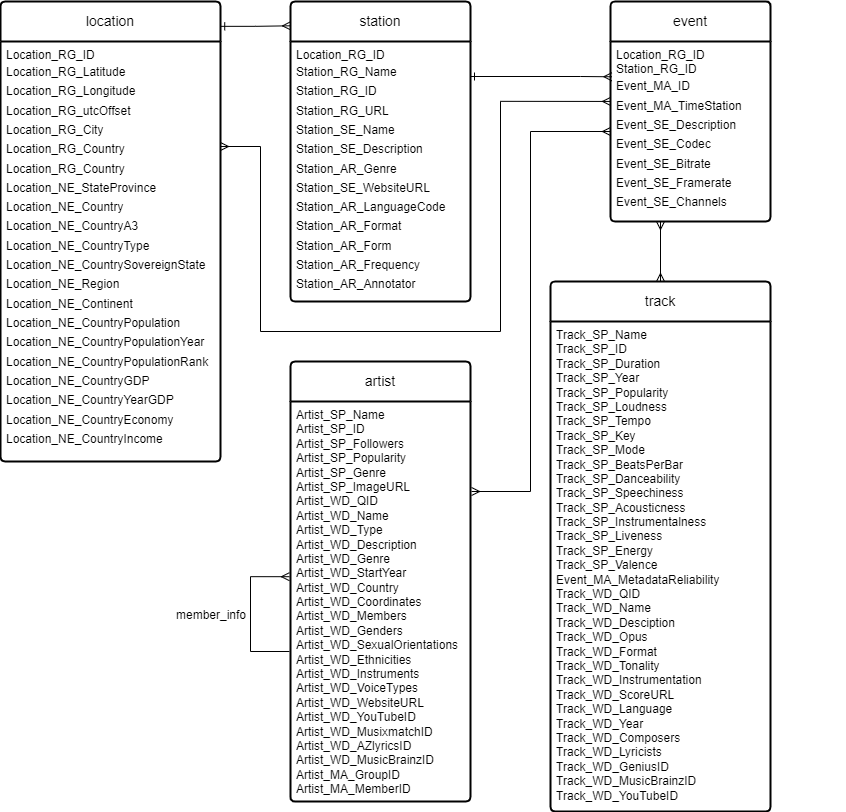}
  \end{minipage}
  \caption{Database tables and connections for the MIRAGE online dashboard.}
  \label{fig:database}
\end{figure}

Shown in Figure~\ref{fig:database}, the database is partitioned into five tables: location, station, event, artist, and track. The data are structured in this manner to facilitate convenient retrieval and examination of each category while minimizing duplication. In this way, the database allows for easy filtering and sorting based on specific criteria, enhancing the overall efficiency of data analysis. Additionally, partitioning of data into separate tables helps to prevent errors and inconsistencies in data entry and manipulation.

\subsection{Layout}\label{subsec:system}
\subsubsection{Earth-View Panel}
The 3D interactive Earth-view (or `globe') panel visualizes the number of stations across the globe. Shown in Figure~\ref{fig:mdash}, each hexagonal-shaped vertical bar identifies the locations where radio stations reside. The height of each bar represents the number of stations at that geographic location, and the bars are also color-coded by country. The Earth-view panel is also linked to the event-list panel such that when a user selects a specific location on the Earth view, the event-list panel automatically filters (i.e., restricts) the station- and event-level metadata to the selected location. In this way, users may compare the number of stations in various regions and discern any recurring patterns or trends.  


\subsubsection{Event-List Panels}
Once users have selected a specific location on the interactive Earth-view panel or using the search function on the toolbox, they can retrieve metadata for the top 1,000 most recent entries in the event-list panel. Users can also select and add events to the selected event-list panel for further analysis and/or export, enabling users to revise their search parameters without losing selected metadata. The event-list panels also allow users to download the contents of either table in CSV format, or obtain a URL to share the results of their most recent search with another user.

\begin{figure}[t!]
  \centering
  \begin{minipage}{1\linewidth} 
    \includegraphics[width=\textwidth,scale=.38]{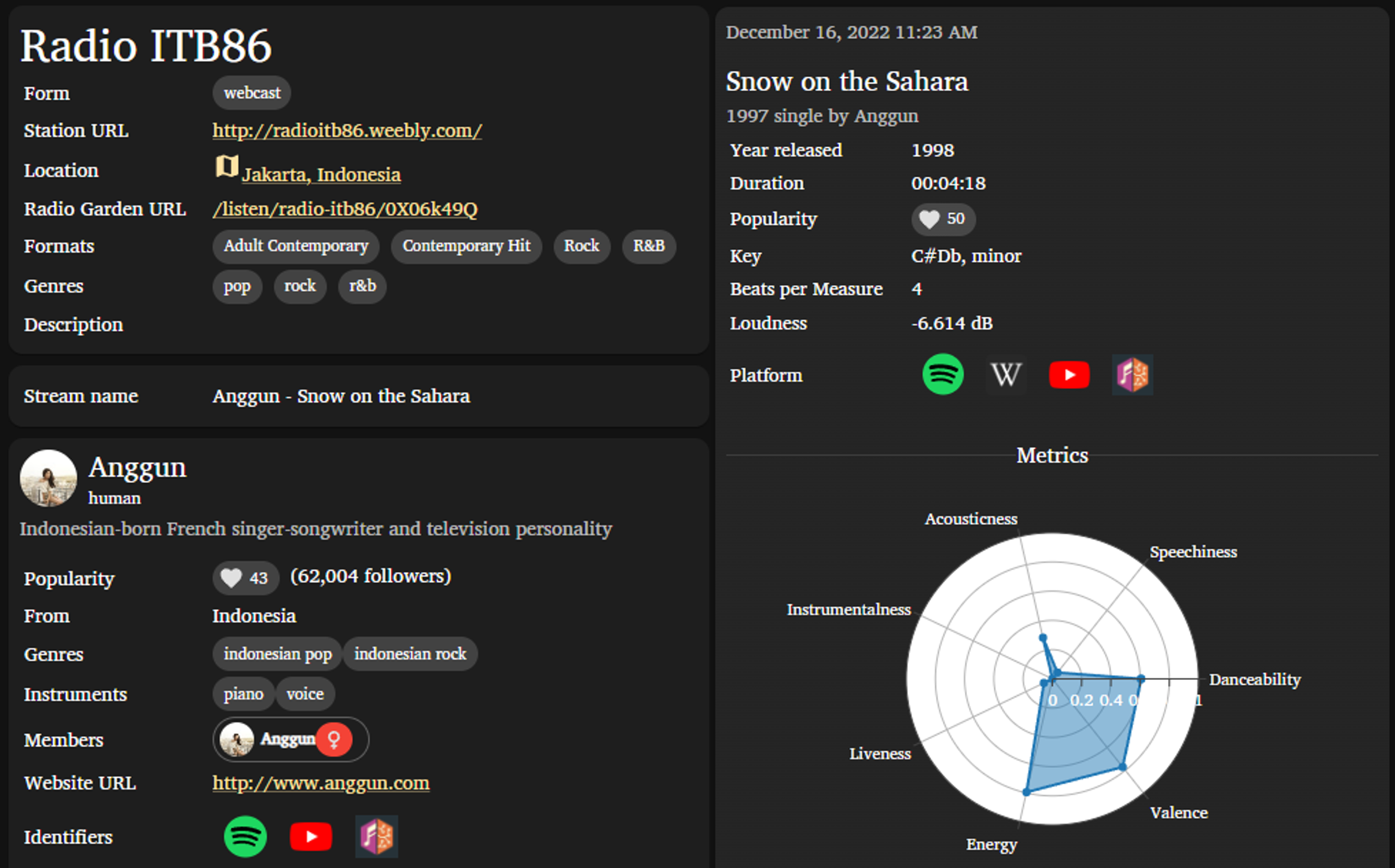}
    \vspace{-.5em}
  \end{minipage}
    \begin{minipage}{1\linewidth} 
    \includegraphics[width=\textwidth,scale=.38]{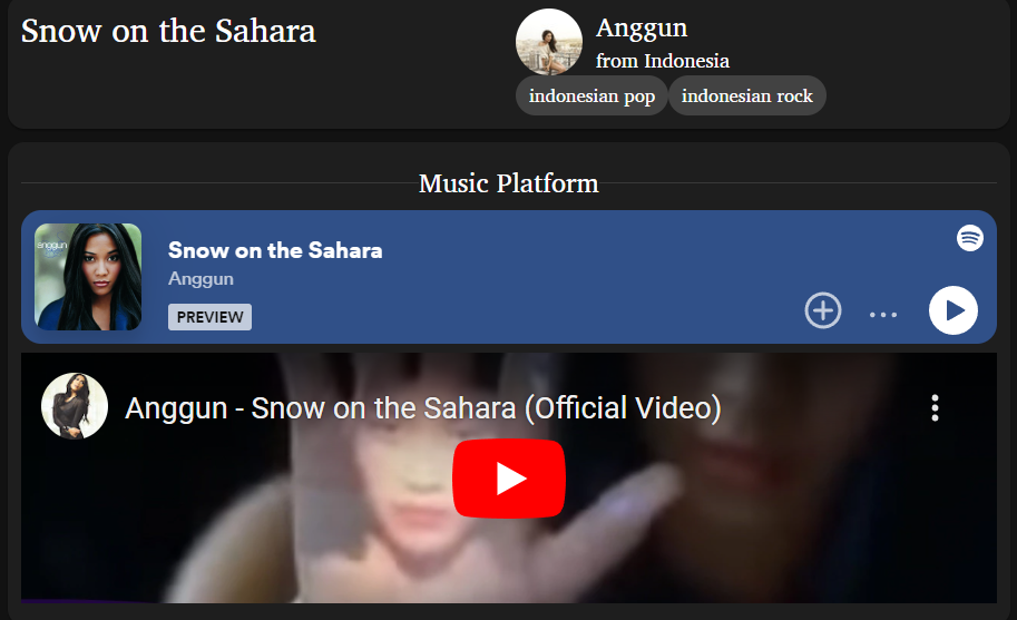}
  \end{minipage}
  \caption{Top: Examples of the event-detail (top) and listen (bottom) panels in the MIRAGE online dashboard.}
  \label{fig:songdetail}
\end{figure}


\begin{figure*}[t!]
  \centering
  \begin{minipage}{1\textwidth} 
    \includegraphics[width=\textwidth,scale=1]{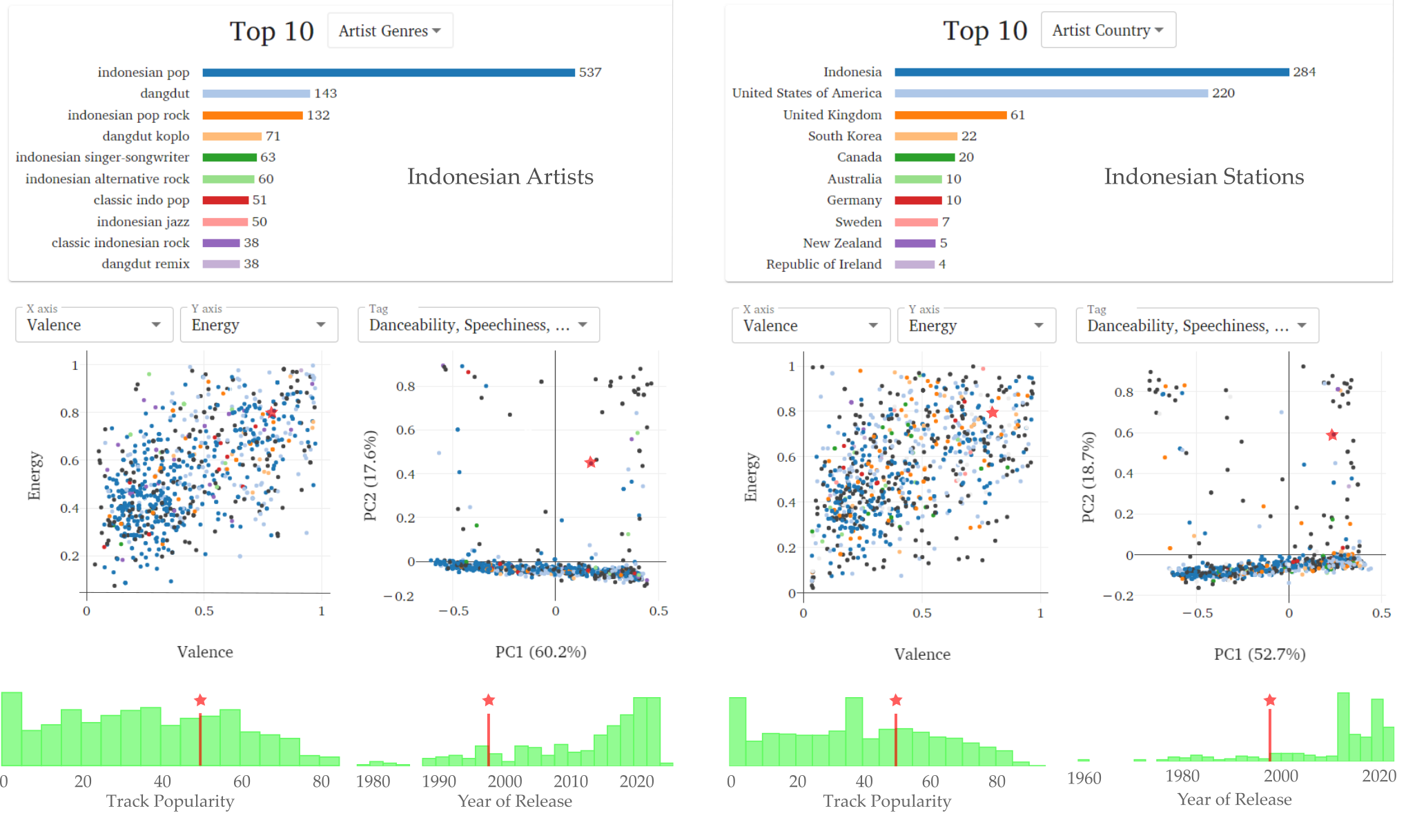}
  \end{minipage}
  \caption{Event-List Visualization Pane for events by Indonesian artists (left) or Indonesian stations (right) in the MIRAGE online dashboard. The red star indicates the position of Anggun's ``Snow on the Sahara.''}
  \label{fig:analysis_fig}
\end{figure*}


\subsubsection{Map Panel}
The map panel enables users to readily visualize the events' geographic distribution in the event-list panel. Each dot on the map reflects the precise position of a particular event, and the size of the dot represents the number of events at that position. If the user selects an event from the event-list panel, that event will be represented by a red dot in the map panel. In this way, the map panel offers users a distinctive method for visualizing the geographical variety of the events in the event-list panel. 


\subsubsection{Event-Detail \& Listen Panels}
Shown in Figure~\ref{fig:songdetail}, the event-detail panel displays the currently selected event from the event-list panel. The content is categorized into four sections: radio-station metadata (e.g., name, location, formats, url, etc.), event metadata (i.e., stream description), artist details (e.g., name(s), gender(s), group affiliations, instrument(s), etc.), and track metadata (e.g., track title, duration, key/mode,  etc.). Figure~\ref{fig:songdetail}, for example, presents all available metadata for Indonesian singer Anggun's ``Snow on the Sahara,'' which streamed on Radio ITB86 in Jakarta, Indonesia on December 16, 2022. Note that users can obtain a list of demographic (e.g., gender, nationality, etc.) and musicological (e.g., list of instruments, vocal type, associated genres, etc.) information about Anggun, review additional metadata about the song itself (year released, language, the song's lyricist(s), etc.), and finally navigate to other websites and online music libraries using the provided hyperlinks.

Finally, the listen panel allows users to stream available recordings using embedded links to the integrated Spotify and YouTube platforms. Although not all events are available on both platforms, the dashboard is regularly updated to ensure that the provided information is current.

\subsubsection{Event-List Visualization Panel}

Shown in Figure~\ref{fig:analysis_fig}, the event-list visualization panel allows users to explore the searched or selected event list using interactive bar, scatter, and histogram plots. For each plot, users may select the appropriate metadata variable(s) from a dropdown list, edit the plot using Plotly Chart Studio,\footnote{\href{chart-studio.plotly.com}{chart-studio.plotly.com}.}, and finally export the plot in SVG format.

\section{Example Use Case}\label{sec: use case}

The metadata and visualizations produced by the MIRAGE online dashboard have numerous applications for users. Figure~\ref{fig:analysis_fig}, for example, examines Anggun's ``Snow on the Sahara'' within the context of events produced by Indonesian artists across the globe (left), or streaming on Indonesian radio stations (right). The MIRAGE-MetaCorpus features 37 events (and 12 tracks) produced by Anggun, of which 21 were ``Snow on the Sahara'' (or its French language version, ``La neige au Sahara''). 

Among Indonesian artists, music genres familiar to western listeners like pop, pop-rock, and alternative rock rank in the top 10, along with characteristic southeast Asian genres like dangdut and koplo. Among events streaming on Indonesian stations, music by Indonesian artists also ranks first, though several Anglophone countries also rank in the top ten (USA, UK, etc.). Scatter plots of a two-dimensional arousal-valence emotion space and a two-component solution from a principal components analysis of the track's danceability, speechiness, acousticness, liveness, and instrumentalness further reveal the track's unconventional expressive and musical characteristics relative to the other tracks produced by Indonesian artists or streaming on Indonesian stations.  Finally, histograms of the track's popularity and year of release reflect the song's enduring popularity more than two decades after its initial release.

\section{Conclusion \& Ethical Considerations}\label{sec:conclusion}
This development release (v0.2) of the MIRAGE online dashboard provides a snapshot of the contemporary global listening landscape for scholars across the (digital) humanities. Our purpose in doing so is to facilitate cross-cultural, comparative research, which has become a pressing concern in several music disciplines \cite{Ewell2020, Jacoby2020, Lacey2018, Savage2012}. To that end, the MIRAGE-MetaCorpus features metadata for 1 million events that streamed on 10,000 radio stations across the globe, and the dashboard is interoperable with several platforms and operating systems \cite{Moss2021, Wilkinson2016}. 

As a metadata repository, the MIRAGE-MetaCorpus contains links to online resources that we do not control. To mitigate the potential for dataset degradation over time, the research team plans to update (and collect additional) metadata annually. Nevertheless, the attribution metadata provided by the radio station's stream encoder does not always reliably match metadata provided by online music libraries. In our view, nonmatching (or `unreliable') metadata allow the research community to evaluate the coverage (i.e., bias) of online music libraries for the music found on internet radio. Nevertheless, MIRAGE users should be aware of the potential for matching errors. For tasks where higher match quality is important, users may search for reliable metadata in the online dashboard, or export reliable subsets of the MIRAGE-MetaCorpus. 

Similarly, this project provides access to metadata and musicological features produced by proprietary (or otherwise undisclosed) algorithms, often trained on western musical traditions and their associated organizational principles. As a result, we encourage the research community to treat the attribution metadata in MIRAGE as a starting point for developing corpora and methodologies involving other musical traditions \cite{Born2020, Huang2023}. 

In developing the MIRAGE online dashboard, the research team has attempted to protect the interests of copyright holders by only including publicly available metadata protected under fair use while enabling users to stream recordings using embedded links to commercial services like Spotify or YouTube. The dashboard also adheres to the user agreements from the libraries and streaming services mentioned above (e.g., Radio Garden, Spotify, WikiData), according to which users may access and interact with all data on the online dashboard, but they may only export public-domain data for further analysis and study (i.e., from the Radio Garden API, the Natural Earth data set, station stream encoder, and WikiData). Perhaps most importantly, this project did not directly record/store audio from station streams at any point in the data-collection pipeline. 

Nevertheless, we acknowledge the concerns of copyright holders (artists, radio stations, online music libraries, and streaming services) who do not wish to share attribution metadata about their work (e.g., artist demographics, track details, etc.). We only provide links to publicly available sources and do not own the copyright for any music referenced in the MIRAGE-MetaCorpus. For that reason, copyright holders may request the removal of metadata from the MIRAGE project.\footnote{Please contact \href{mailto:miragedashboard@gmail.com}{miragedashboard@gmail.com}.}  

In addition to completing station-list review for the remaining stations in MIRAGE-MetaCorpus, future versions of the dashboard will transition from React+Nodejs to Remix in order to enhance the speed of queries and allow users to access and review more than 1,000 events simultaneously in the event-list panel. The team also plans to conduct a usability study to examine the dashboard's practical utility, as well as incorporate additional customizable sampling and visualization tools like statistical surface maps to enhance the user's exploration of metadata variables in MIRAGE \cite{OSullivan2010}. In doing so, we hope future versions of this dashboard will facilitate cross-cultural, comparative research for a medium that places diversity center stage. 

\bibliography{ISMIRtemplate}

\begin{thebibliography}{10}
\providecommand{\url}[1]{#1}
\csname url@samestyle\endcsname
\providecommand{\newblock}{\relax}
\providecommand{\bibinfo}[2]{#2}
\providecommand{\BIBentrySTDinterwordspacing}{\spaceskip=0pt\relax}
\providecommand{\BIBentryALTinterwordstretchfactor}{4}
\providecommand{\BIBentryALTinterwordspacing}{\spaceskip=\fontdimen2\font plus
\BIBentryALTinterwordstretchfactor\fontdimen3\font minus \fontdimen4\font\relax}
\providecommand{\BIBforeignlanguage}[2]{{%
\expandafter\ifx\csname l@#1\endcsname\relax
\typeout{** WARNING: IEEEtran.bst: No hyphenation pattern has been}%
\typeout{** loaded for the language `#1'. Using the pattern for}%
\typeout{** the default language instead.}%
\else
\language=\csname l@#1\endcsname
\fi
#2}}
\providecommand{\BIBdecl}{\relax}
\BIBdecl

\bibitem{Lacey2018}
K.~Lacey, ``Up in the air? the matter of radio studies,'' \emph{Radio Journal: International Studies in Broadcast \& Audio Media}, vol.~16, no.~2, pp. 109--126, 2018.

\bibitem{Bottomley2020}
A.~J. Bottomley, \emph{Sound streams: A cultural history of radio-internet convergence}.\hskip 1em plus 0.5em minus 0.4em\relax Ann Arbor, MI: University of Michigan Press, 2020.

\bibitem{Glantz2016}
M.~Glantz, ``Internet radio adopts a human touch: A study of 12 streaming music services,'' \emph{Journal of Radio \& Audio Media}, vol.~23, no.~1, p. 36–49, 2016.

\bibitem{Wall2004}
T.~Wall, ``The political economy of internet music radio,'' \emph{The Radio Journal}, vol.~2, no.~1, p. 27–44, 2004.

\bibitem{Chambers2003}
T.~Chambers, ``Radio programming diversity in the era of consolidation,'' \emph{Journal of Radio Studies}, vol.~10, no.~1, p. 33–45, 2003.

\bibitem{Hendy2000}
D.~Hendy, \emph{Radio in the global age}.\hskip 1em plus 0.5em minus 0.4em\relax Cambridge, UK: Polity Press, 2000.

\bibitem{Uimonen2017}
H.~Uimonen, ``Beyond the playlist: Commercial radio as music culture,'' \emph{Popular Music}, vol.~36, no.~2, pp. 178--195, 2017.

\bibitem{Burgoyne2011}
J.~A. Burgoyne, J.~Wild, and I.~Fujinaga, ``An expert ground-truth set for audio chord recognition and music analysis,'' in \emph{Proceedings of the 12th International Society for Music Information Retrieval Conference (ISMIR)}, A.~Klapuri and C.~Leider, Eds., Miami, FL, 2011, pp. 423--428.

\bibitem{Declercq2011}
T.~de~Clercq and D.~Temperley, ``A corpus analysis of rock harmony,'' \emph{Popular Music}, vol.~30, pp. 47--70, 2011.

\bibitem{Sears2021}
D.~R.~W. Sears and D.~Forrest, ``Triadic patterns across classical and popular music corpora: Stylistic conventions, or characteristic idioms?'' \emph{Journal of Mathematics and Music}, vol.~15, no.~2, p. 140–153, 2021.

\bibitem{Bertin-Mahieux2011}
T.~Bertin-Mahieux, D.~P.~W. Ellis, B.~Whitman, and P.~Lamere, \emph{The million song dataset}, Miami, FL, 2011.

\bibitem{Aizenberg2012}
N.~Aizenberg, Y.~Koren, and O.~Somekh, \emph{Build your own music recommender by modeling internet radio streams}, Lyon, France, 2012.

\bibitem{Dang2012}
T.~Dang, A.~Anand, and L.~Wilkinson, \emph{FmFinder: Search and filter your favorite songs}.\hskip 1em plus 0.5em minus 0.4em\relax Heidelberg, Germany: Springer-Verlag, 2012, vol. 7431.

\bibitem{Silva2017}
G.~R.~L. Silva, L.~M. de~Oliveira, R.~R. de~Medeiros, O.~Goussevskaia, and F.~Benevenuto, ``Characterizing internet radio stations at scale,'' in \emph{Proceedings of the International Conference on Web Intelligence (WI 2017)}.\hskip 1em plus 0.5em minus 0.4em\relax Association for Computing Machinery, 2017, Conference Proceedings, pp. 670--677.

\bibitem{AlaFossi2008}
M.~Ala-Fossi, S.~Lax, B.~O'Neill, and H.~Shaw, ``The future of radio is still digital--but which one? expert perspectives and future scenarios for radio media in 2015,'' \emph{Journal of Radio \& Audio Media}, vol.~15, no.~1, p. 4–25, 2008.

\bibitem{Lind1999}
R.~A. Lind and N.~J. Medoff, ``Radio stations and the world wide web,'' \emph{Journal of Radio \& Audio Media}, vol.~6, no.~2, pp. 203--221, 1999.

\bibitem{Kuhn2011}
F.~Kuhn, ``Internet radio flows: Between the local and the global,'' \emph{Radio Journal: International Studies in Broadcast \& Audio Media}, vol.~9, no.~1, pp. 35--49, 2011.

\bibitem{Sears2024}
\BIBentryALTinterwordspacing
D.~R.~W. Sears, ``{Music Informatics for Radio Across the Globe (MIRAGE) MetaCorpus (v0.2)},'' Jul. 2024. [Online]. Available: \url{https://doi.org/10.5281/zenodo.12786202}
\BIBentrySTDinterwordspacing

\bibitem{Ewell2020}
\BIBentryALTinterwordspacing
P.~A. Ewell, ``Music theory and the white racial frame,'' \emph{Music Theory Online}, vol.~26, no.~2, 2020. [Online]. Available: \url{https://www.mtosmt.org/issues/mto.20.26.2.ewell.php}
\BIBentrySTDinterwordspacing

\bibitem{Jacoby2020}
N.~Jacoby, E.~H. Margulis, M.~Clayton, E.~Hannon, H.~Honing, J.~Iversen, T.~R. Klein, S.~A. Mehr, L.~Pearson, I.~Peretz, M.~Perlman, R.~Polak, A.~Ravignani, P.~E. Savage, G.~Steingo, C.~J. Stevens, L.~J. Trainor, S.~E. Trehub, and M.~Veal, ``Cross-cultural work in music cognition: Challenges, insights, and recommendations,'' \emph{Music Perception}, vol.~37, no.~3, pp. 185--195, 2020.

\bibitem{Savage2012}
P.~E. Savage and S.~Brown, ``Toward a new comparative musicology,'' \emph{Analytical Approaches to World Music}, vol.~2, no.~2, pp. 149 -- 197, 2013.

\bibitem{Moss2021}
F.~C. Moss and M.~Neuwirth, ``Fair, open, linked: Introducing the special issue on open science in musicology,'' \emph{Empirical Musicology Review}, vol.~16, no.~1, pp. 1--4, 2021.

\bibitem{Wilkinson2016}
M.~D. Wilkinson, M.~Dumontier, I.~J. Aalbersberg, ..., and B.~Mons, ``The fair guiding principles for scientific data management and stewardship,'' \emph{Scientific Data}, vol.~3, no.~1, p. 160018, 2016.

\bibitem{Born2020}
G.~Born, ``Diversifying {MIR}: Knowledge and real-world challenges, and new interdisciplinary futures,'' \emph{Transactions of the International Society for Music Information Retrieval}, vol.~3, no.~1, pp. 193--204, 2020.

\bibitem{Huang2023}
R.~S. Huang, A.~Holzapfel, B.~L. Sturm, and A.~K. Kaila, ``Beyond diverse datasets: Responsible {MIR}, interdisciplinarity, and the fractured worlds of music,'' \emph{Transactions of the International Society for Music Information Retrieval}, vol.~6, no.~1, pp. 43--59, 2023.

\bibitem{OSullivan2010}
D.~O'Sullivan and D.~J. Unwin, \emph{Geographic Information Analysis}, 2nd~ed.\hskip 1em plus 0.5em minus 0.4em\relax Hoboken, NJ: Wiley, 2010.

\end{thebibliography}

\end{document}